\documentclass{article}

\usepackage{microtype}
\usepackage{graphicx}
\usepackage{subfigure}
\usepackage{booktabs} 

\usepackage{hyperref}

\usepackage[accepted]{icml2024}

\usepackage{amsmath}
\usepackage{amssymb}
\usepackage{mathtools}
\usepackage{amsthm}

\usepackage[capitalize,noabbrev]{cleveref}

\theoremstyle{plain}

\theoremstyle{definition}

\theoremstyle{remark}

\usepackage[textsize=tiny]{todonotes}

\usepackage{float}
\usepackage{amsmath}
\usepackage{xcolor}
\usepackage{enumitem}
\usepackage{cleveref}
\usepackage{caption}
\usepackage{subcaption}
\usepackage{overpic}
\usepackage{listings}
\usepackage{color}
\usepackage{booktabs}
\usepackage{mathtools}
\usepackage{amsfonts}
\usepackage{array}

\newcommand{\ie}{\textit{i.e.}}
\newcommand{\eg}{\textit{e.g.}}

\lstdefinestyle{json}{
    basicstyle=\tiny\ttfamily, 
    numberstyle=\scriptsize,
    numbersep=8pt,
    showstringspaces=false,
    breaklines=true,
    frame=lines,
    backgroundcolor=\color{white},
    stringstyle=\color{cyan},
    keywordstyle=\color{blue},
    commentstyle=\color{olive},
    literate=
     *{0}{{{\color{red}0}}}1
      {1}{{{\color{red}1}}}1
      {2}{{{\color{red}2}}}1
      {3}{{{\color{red}3}}}1
      {4}{{{\color{red}4}}}1
      {5}{{{\color{red}5}}}1
      {6}{{{\color{red}6}}}1
      {7}{{{\color{red}7}}}1
      {8}{{{\color{red}8}}}1
      {9}{{{\color{red}9}}}1
      {:}{{{\color{blue}:}}}1
      {,}{{{\color{blue},}}}1
      {\{}{{{\color{blue}\{}}}1
      {\}}{{{\color{blue}\}}}}1
      {[}{{{\color{blue}[}}}1
      {]}{{{\color{blue}]}}}1
}

\definecolor{background}{gray}{0.95}

\icmltitlerunning{Efficient Generation of Multimodal Fluid Simulation Data}

\begin{document}

\twocolumn[
\icmltitle{Efficient Generation of Multimodal Fluid Simulation Data}





\begin{icmlauthorlist}
\icmlauthor{Daniele Baieri}{sap}
\icmlauthor{Donato Crisostomi}{sap}
\icmlauthor{Stefano Esposito}{avg}
\icmlauthor{Filippo Maggioli}{mb}
\icmlauthor{Emanuele Rodol\`a}{sap}
\end{icmlauthorlist}

\icmlaffiliation{sap}{Department of Computer Science, Sapienza University of Rome, Italy}
\icmlaffiliation{avg}{Autonomous Vision Group, University of Tuebingen, Germany}
\icmlaffiliation{mb}{Department of Computer Science, University of Milano-Bicocca, Italy}

\icmlcorrespondingauthor{Daniele Baieri}{baieri@di.uniroma1.it}


\vskip 0.3in
    ]



\printAffiliationsAndNotice{\icmlEqualContribution} 

\begin{abstract}
In this work, we introduce an efficient generation procedure to produce synthetic multi-modal datasets of fluid simulations. The procedure can reproduce the dynamics of fluid flows and allows for exploring and learning various properties of their complex behavior, from distinct perspectives and modalities. We employ our framework to generate a set of thoughtfully designed training datasets, which attempt to span specific fluid simulation scenarios in a meaningful way. 
The properties of our contributions are demonstrated by evaluating recently published algorithms for the neural fluid simulation and fluid inverse rendering tasks using our benchmark datasets. 
Our contribution aims to fulfill the community's need for standardized training data, fostering more reproducibile and robust research. 
\end{abstract}

\section{Introduction}
\label{sec:introduction}

Applying the representational power of machine learning to the prediction of complex fluid dynamics has been a relevant subject of study for years and is regarded today as an established research field~\cite{lino:2023:dlfluidsstar}.
However, the amount of available fluid simulation data does not match the notoriously high requirements of machine learning methods. Researchers have typically addressed this issue by generating their own datasets, as in the case of \citet{pfaff2021learning} and \citet{Ummenhofer2020Lagrangian}, only to cite a few. Such an heterogeneous distribution of training data across different methods prevents a consistent comparison of the proposed approaches.

This work introduces a generation procedure for synthetic multi-modal fluid simulations datasets. By leveraging a GPU implementation, our procedure is efficient enough that no data needs to be exchanged between users, except for configuration files required to reproduce the dataset. Furthermore, our procedure allows multiple modalities (generating both geometries and photorealistic renderings) and is general enough for it to be applied to various tasks in data-driven fluid simulation. 


\begin{figure}[t]
    \centering
    \includegraphics[width=\linewidth]{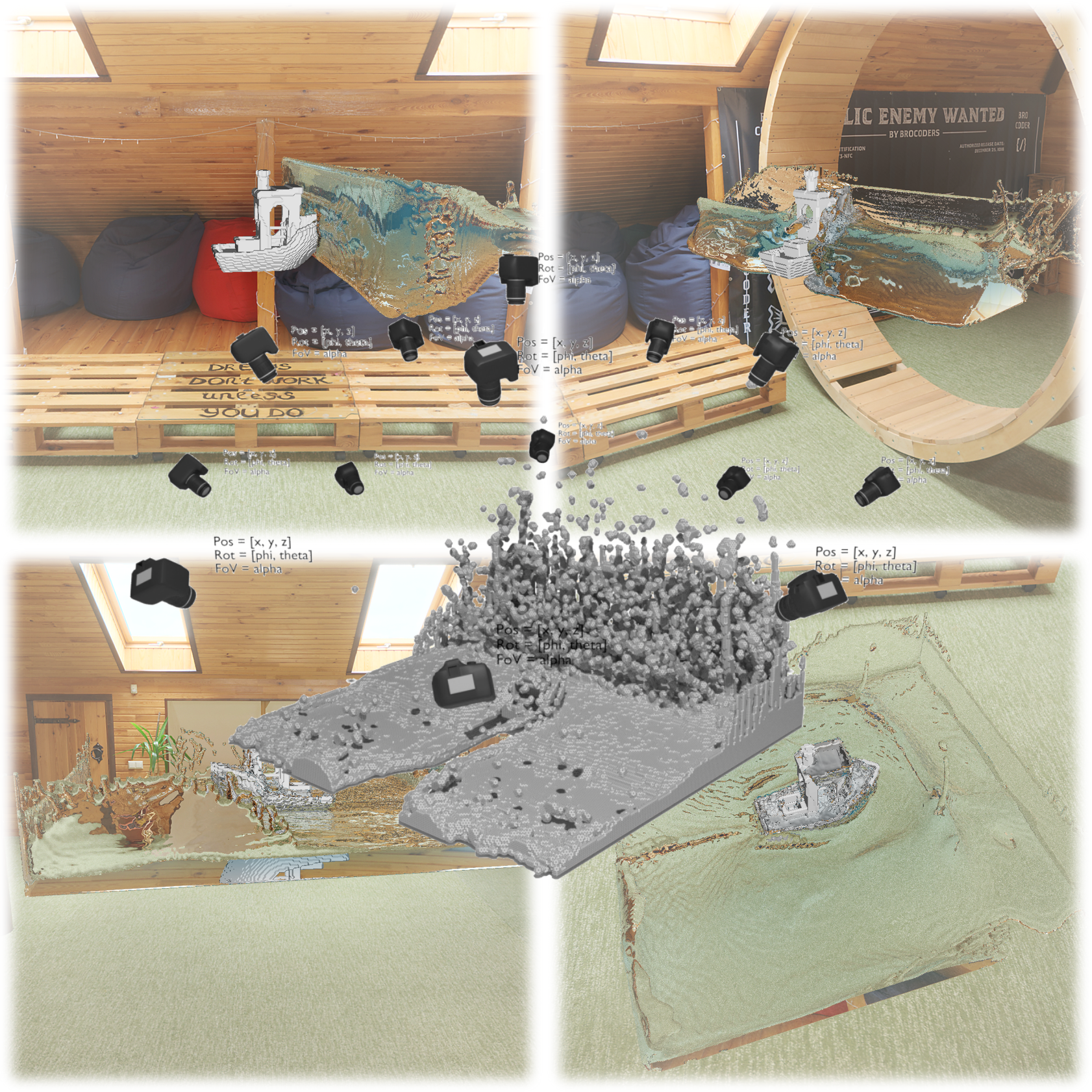}
    
    
    \caption{Our generation tool performs Lattice Boltzmann simulation steps, exporting the fluid geometry and rendering output frames from multiple viewpoints. Camera positions are sampled with Fibonacci sphere sampling~\cite{keinert:2015:fibonaccisphere}, and exported as metadata. 
    }
    \label{fig:nerf-cameras}
\end{figure}

Recent advancements in animation datasets have significantly enhanced data-driven models, streamlining traditional computer graphics challenges and expanding digital artists' creative possibilities. Notably, skeleton-based animation datasets, supported by specialized data capture technology, have become prevalent due to their diverse applications. Prominent examples include DFAUST~\cite{dfaust} for human body animations and DeformingThings4D~\cite{deformingthings} for various subjects. One really successful application of human motion datasets was the creation of SMPL \cite{smpl}, a skinned multi-person linear model, which may be used for traditional skinning as well as in a generative fashion, to sample novel human poses and styles. However, capturing intricate fluid deformations contrasts sharply, necessitating specialized, expensive, and labor-intensive tools~\cite{hoyer:2005:3dscanfluids, scalarflow}. This complexity restricts the diversity of fluid motion data, prompting the creation of synthetic datasets to encompass a broader spectrum of fluid behaviors.

Our dataset is constructed using a flexible template that is customizable through a configuration file, defining variables and constants such as initial fluid states, boundary conditions, and simulation parameters. The template is expandable to incorporate extra features like initial velocities or dynamic force fields. Scenes are generated by varying these parameters and simulated using a highly efficient GPU-based Lattice Boltzmann method~\cite{fluidx3d}, enabling rapid large-scale dataset creation. Additionally, users can select multiple viewpoints for photorealistic renders and choose the fluid geometry's export format, accommodating both Eulerian and Lagrangian simulation data.

Our dataset holds potential for various research areas. We demonstrate its utility in two key domains that have garnered considerable attention: advancing data-driven fluid simulation models~\cite{Ummenhofer2020Lagrangian,fluidsuperresgan} and tackling inverse rendering/surface recovery~\cite{li2023pacnerf,chu2022physics}.

Overall, our work aims to bridge a long-standing gap in data-driven fluid simulation research, enhancing community contributions, reproducibility, and systematic evaluation. Our key contributions are:
\begin{itemize}
    \item introducing an efficient GPU-based framework for generating synthetic, multi-modal fluid simulation data, capturing diverse fluid dynamics representations (see \Cref{fig:nerf-cameras});
    \item using this framework, we provide three training datasets for distinct fluid simulation scenarios, setting a benchmark for consistent research assessments;
    \item demonstrating the effectiveness of our datasets by successfully training state-of-the-art models for fluid simulation and inverse rendering tasks.
\end{itemize}

\section{Motivation and applications}


Our dataset could substantially aid efforts integrating neural networks with fluid dynamics simulation, offering new research avenues. Notably, it supports fluid super-resolution, where models learn to enhance low-resolution simulations into detailed, high-resolution outputs. GAN models have been notably effective in this area~\cite{fluidsuperresgan,xie2018tempoGAN}, improving super-resolution in scenarios like smoke flow upsampling and turbulence prediction~\cite{bai:2020:upsamplingsmoke,bai:2021:predictingturbulence}. Further research extends this approach to complex scenarios like multi-phase flows~\cite{li2022using} and label-free super-resolution~\cite{gao2021super}. This methodology also shows promise in simulating biological systems, indicating its broad applicability~\cite{superresbiology}.

More in general, one could exploit neural networks to learn the entire simulation model using the current fluid state as input, adopting either Lagrangian or Eulerian perspectives. The Lagrangian approach focuses on particle-based fluid dynamics. Continuous convolution on point sets has been shown to yield efficient and robust models~\cite{Ummenhofer2020Lagrangian}. Enhancements include momentum conservation~\cite{prantl2022dmcf} and applying graph neural networks to particle-based models~\cite{li:2022:neuralnetfluid}, improving performance without sacrificing quality. Conversely, the Eulerian viewpoint treats fluids and properties (like velocity) as functions over space. 
Innovations here include generative models for velocity fields~\cite{deepfluids} and using latent spaces in generative models to produce stable and controllable simulations~\citet{wiewel2019latent,wiewel2003latent}. We refer to \citet{vinuesa:2022:fluidlearning} for a comprehensive review of learning paradigms in fluid simulation.

More recent research explores NeRF's inverse rendering technique~\cite{mildenhall2020nerf} in fluid dynamics. \citet{chu2022physics} first integrated Navier-Stokes principles into 4D NeRF training, developing a dual-network model enforcing physical constraints to accurately reconstruct smoke density with limited camera views. 
NeuroFluid~\cite{guan2022neurofluid} built on this, adopting a Lagrangian approach and incorporating physics constraints into volume rendering for model optimization. \citet{li2023pacnerf} combined Eulerian and Lagrangian methods, achieving remarkable versatility across diverse materials. However, these approaches overlook real-world fluid material properties, treating color as a mere surface texture. NeReF~\cite{Wang2022NeReFNR} addressed this by formulating a NeRF that considers reflective and transmissive materials, enabling realistic light-fluid interactions, notably in transparent fluids like water.

A prevalent problem in the cited research is the absence of a standardized benchmark dataset, leading researchers to create new simulation datasets for method evaluation. Our generation tool addresses this by enabling: a) export of fluids as both particles and density fields, b) multi-view renderings of simulations, and c) efficient dataset regeneration using a configuration file, eliminating the need to distribute massive data volumes.

\section{Related work}

Our dataset, while not the first in fluid simulation data, differs significantly from prior works like the Johns Hopkins Turbulence Databases (JHTD)~\cite{jhtd2007, jhtd2008,graham2016turbulent}. JHTD offers extensive, high-resolution data on turbulent flows, targeting geophysical and engineering applications like meteorology and aerodynamics. However, its massive scale and detail, with thousands of frames and billions of voxels per frame, render it less suitable for computer graphics and vision applications. In contrast, our contribution is designed with practicality and adaptability for CG/CV purposes.

%

\begin{figure}[h]
    \centering
    \includegraphics[width=\linewidth]{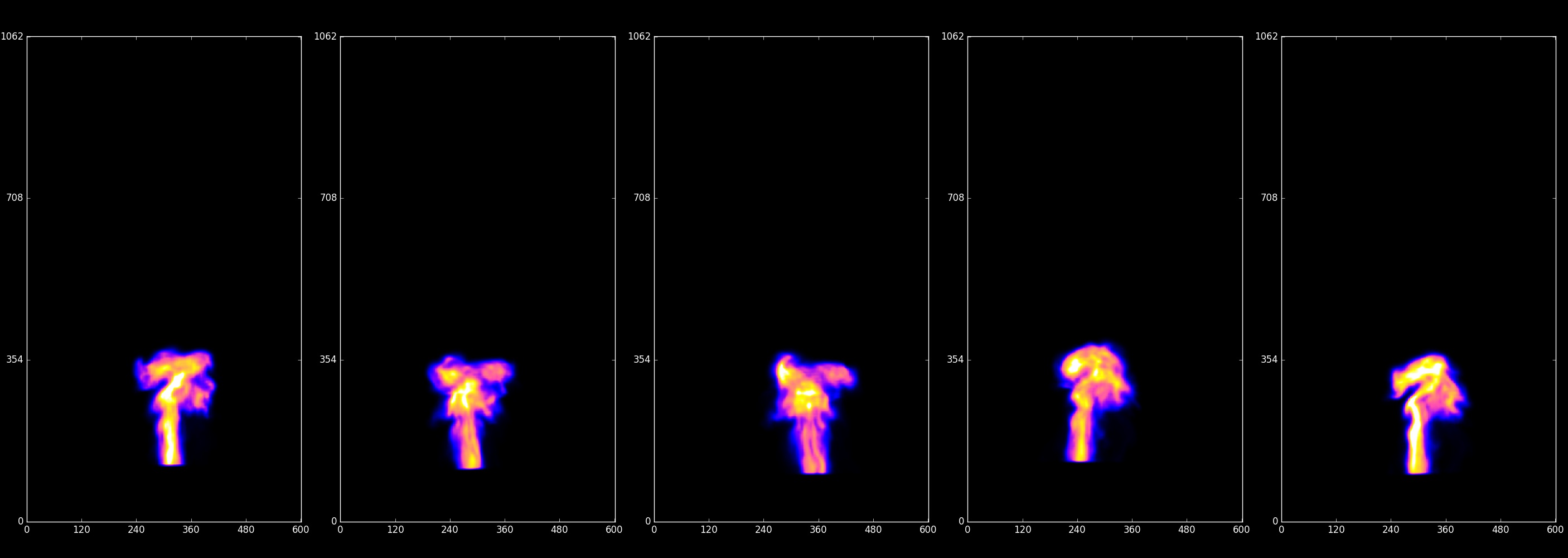}
    \includegraphics[width=\linewidth]{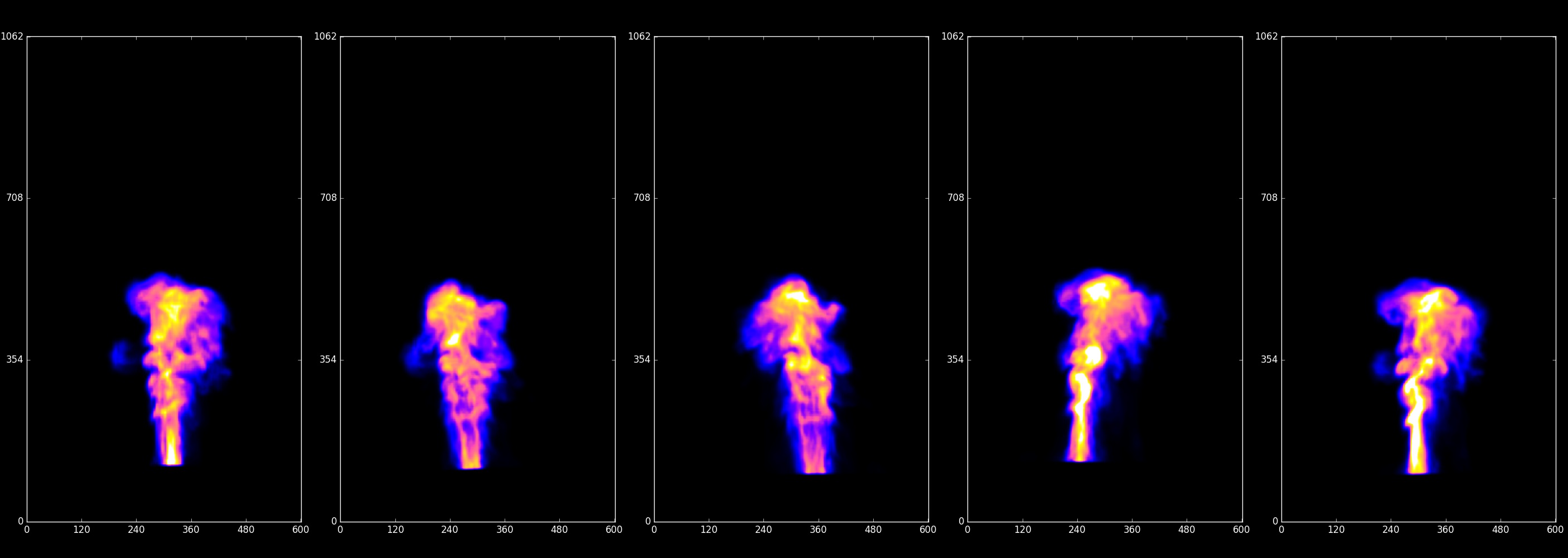}
    \includegraphics[width=\linewidth]{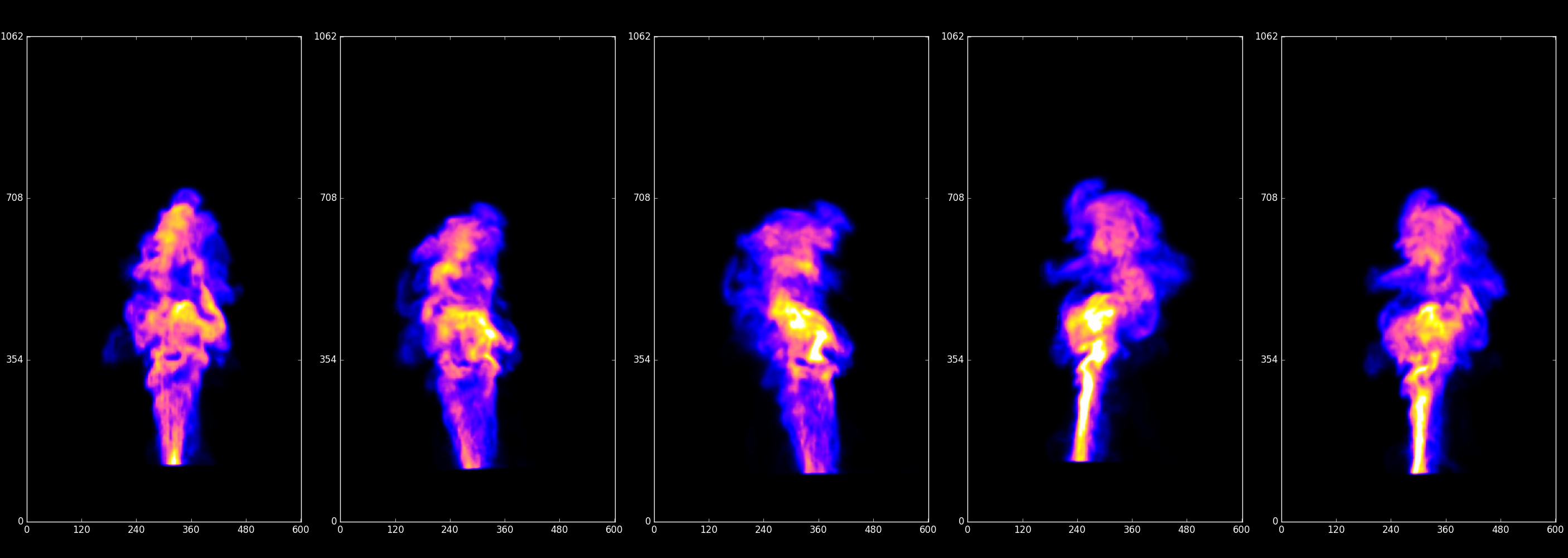}
    \caption{Multiple rendered views for three timesteps of a smoke plume simulation from the ScalarFlow~\cite{scalarflow} dataset. The rendered density field is reconstructed from real smoke captures.}
    \label{fig:scalarflow}
\end{figure}

Many datasets accompanying data-driven simulation studies have limitations. \citet{pfaff2021learning}'s work, using graph neural networks~\cite{scarselli2008graph} on triangle meshes, was confined to 2D due to planar mesh limitations in representing 3D volumes. EAGLE dataset~\cite{janny2023eagle}, also mesh-based, focused on turbulent wind dynamics influenced by boundary shapes. \citet{stachenfeld2022learned} introduced volumetric datasets across different dimensions, modeling real-world chaotic systems. However, our dataset is specifically tailored for fluid dynamics in computer graphics, offering additional visual data. Notably, datasets from \citet{Ummenhofer2020Lagrangian}, initially minor contributions, later became foundational for training methods like those in \citet{prantl2022dmcf}.


The work most similar in spirit to ours is ScalarFlow \cite{scalarflow}. This is a collection of video captures of real-world smoke plumes dynamics (see \Cref{fig:scalarflow}), coupled with density field reconstructions of the corresponding volumetric data computed by an algorithm proposed by the authors. Despite ScalarFlow being a large scale dataset (ca. 26 billion voxels of data), its scope is limited to the real-world capture setting set up by the authors; when using real data over synthetic would provide no significant advantage, our generation framework could provide similar data in arbitrary quantities, without the additional cost of capture.

Recent proposals by \citet{toshev2023lagrangebench, toshev_2023_10021926} provided contributions in similar,  albeit complementary, directions to our own: their LagrangeBench tool and data allow to sistematically and reproducibly test pre-trained neural Lagrangian simulations models on small synthetic benchmark datasets consisting of well-known simulation settings. Such efforts further highlight the demand for such resources in the field. Combining these efforts could create a comprehensive pipeline for training and evaluating neural Lagrangian simulation models.
Lastly, we mention that large-scale simulation data is widely used beyond fluid dynamics, notably in garment animation. Two notable examples from this field are the datasets presented in CLOTH3D \cite{bertiche2020cloth3d} and TailorNet \cite{patel20tailornet}.

\section{Dataset}

In this section, we outline our data generation process, providing high-level implementation details. For comprehensive documentation, please refer to our code release at \url{see-anonymized-supplementary-material}. We use this procedure to create two mid-scale training datasets for data-driven fluid simulation and a set of scenes for fluid inverse rendering algorithms.


\subsection{Data generation}\label{sec:generation}


Our C++ code library refactors the FluidX3D Lattice-Boltzmann GPU simulator \cite{fluidx3d}. This was chosen considering its a) impressive efficiency on mid-range hardware, b) built-in ray-traced rendering capabilities, and c) wide compatibility with several operating systems and GPU architectures due to its OpenCL kernel implementation. We retained the core implementation and improved the API to expose most functionalities via a generic \texttt{Scene} class. These objects represent the simulation's initial state and parameters, facilitating fine-grained configuration and inheritance for extending base behavior. For instance, initial fluid shapes and boundary conditions can be provided through input mesh files, which are adjustable in translation, rotation, and scaling. Enabling specialized features like point-wise external force fields or inflows/outflows requires overriding basic \texttt{Scene} functionalities.

\begin{lstlisting}[float, style=json, caption={
JSON configuration file for our generation procedure. Variables are sampled and
aggregated in order to generate configuration files for individual instances in the dataset.
Fluid bodies and solid obstacles can be specified both as constants and variables, 
and sampled according to specific properties of geometric primitives or meshes (\eg, center, scale, rotation).
}]
"export_root": "path/to/out/dir",
"seed": 123456,
"constants": {
    "sim_params": {
        "Nx": 256, "Ny": 256, "Nz": 256,
        ...
    },
    "obstacles": {
        ...
    },
    ...
},
"variables": {
    "sim_params": {
        "nu": {
            "type": "linscale",
            "vmin": 0.0005, "vmax": 0.005, "steps": 20
        },
        "sigma": {
            "type": "normal",
            "mean": 0.001, "std": 0.0005, "nsamples": 20
        }
    },
    "fluids": [
        ...
    ]
}
\end{lstlisting}

The refactored implementation serves as a library for defining dataset-specific \texttt{Scene} subclasses, which can be accessed through an executable file. These subclasses are entirely configurable using a simple JSON file (see Listing 1). Using this framework, we implemented the generation procedure as a Python script that creates configuration files and feeds them to the simulation tool. The script takes various input parameters, including global generation parameters like the seed for reproducibility. It also requires a set of constants and variables. Constants are preserved in generated instances, while the variables are sampled based on their selected type, which includes linear intervals, uniform distributions, normal distributions, collections, and more complex structures. Users can also set a maximum scene count, in which case scene instances are uniformly sampled from all possible combinations of variables. This configuration is easily supplied via a separate JSON file.

During simulation, specific frames can be extracted both as multi-view renderings and geometric information. For rendering efficiency, we employ a specialized FluidX3D algorithm that balances speed over photorealism. It limits camera ray bounces to two and uses the mean of material BSDF, rather than sampling them; this way, the light interactions of the water material we use for fluids can still be displayed, while the rendering procedure remains really fast, being executed on the GPU (see \Cref{sec:generationperformance}). For geometric data extraction, we offer two options. The first exports fluid as a binary particle array, with the number of particles determined by the configuration file input. The second exports the fluid's density field as a memory-efficient 3D matrix in sparse coordinate format (COO), stored as a 1D binary array.  Despite the latter method being very memory-efficient (see \Cref{sec:generationperformance}), selecting a specific number of particles may be more convenient at high resolutions. We provide a Python loader to Numpy arrays for both file formats.



\subsection{Datasets}

\subsubsection{Coherent subsets of simulation space}\label{sec:benchmarks}

The first application we present for the procedure we described in \Cref{sec:generation} is devoted to the generation of collections of ``similar'' (in a fluid dynamics sense) simulations, regardless of the scale of such collections. To this end, we introduce two medium scale benchmark datasets based on well-known fluid dynamics settings. Throughout this section, we will refer to 3D coordinate systems assuming that the up-axis is the $z$ direction.

Our first training dataset (\texttt{obstacles}) consists of simulations displaying a sphere of fluid colliding with varying boundary geometries under the effect of gravity. 
We define a cubic boundary $B = (\mathbf{p}, \mathbf{q})$ enveloping the entire scene, where $\mathbf{p}_{\{x,y,z\}}$ and  $\mathbf{q}_{\{x,y,z\}}$ are the min/max coordinate values for the vertices of $B$, respectively. 
We represent the initial fluid state by a sphere $S=(\mathbf{c}, r)$ and the solid obstacle as a triangle mesh $M=(V,F)$, which we can instantiate by sampling its center, rotation and size $(\mathbf{o}, \mathbf{\theta}, s)$. In order to ensure interaction between the two, we set $\mathbf{c}_x=\mathbf{o}_x={(\mathbf{q}_x - \mathbf{p}_x)}/{2}$ and $\mathbf{c}_y=\mathbf{o}_y={(\mathbf{q}_y - \mathbf{p}_y)}/{2}$. 
We sample the radius $r\sim\mathcal{U}(r_{\text{min}}, r_{\text{max}})$ and mesh size $s\sim\mathcal{U}(s_{\text{min}}, s_{\text{max}})$ as the maximum dimension of its bounding box. Since we want the fluid above the obstacle to ensure interaction, we sample the two in the top/bottom halves of the simulation domain, respectively.  Thus, we sample:
\begin{align} 
    \mathbf{c}_z&\sim \mathcal{U}\left(\dfrac{\mathbf{q}_z - \mathbf{p}_z}{2}+r_{\text{max}},\mathbf{q}_z-r_{\text{max}}\right) \\
    \mathbf{o}_z&\sim \mathcal{U}\left(\mathbf{p}_z + \dfrac{s_{\text{max}}}{2}, \dfrac{\mathbf{q}_z - \mathbf{p}_z}{2} - \dfrac{s_{\text{max}}}{2}\right)\,,
\end{align}
where we choose $r_{\text{min}}, r_{\text{max}}, s_{\text{min}},$ and $s_{\text{max}}$ to be, respectively, the $10\%, 20\%, 25\%$ and $45\%$ of $\mathbf{q}_z - \mathbf{p}_z$.
Lastly, we randomly rotate $M$ along $x$ to augment the dataset with additional unique boundary interactions, \ie, $\mathbf{\theta}_x\sim\mathcal{U}(0, \frac{\pi}{2})$. The geometry of the mesh $(V, F)$ is chosen randomly from a small collection of 5 shapes with varying complexity.

The second dataset, which we refer to as \texttt{dam-break}, is a collection of instances of this well-known fluid dynamics scenario. In this setting, a cuboid of fluid is placed in contact with the lower face of a cubic boundary $B = (\mathbf{p}, \mathbf{q})$, and simulated under the effect of gravity. The only variable for our scene generation procedure is the initial placement of the dam inside the scene: if we represent the cuboid by its center and sides, \ie, $D = (\mathbf{c}, \mathbf{s})$, we set the $y$ coordinates to span the entire $y$-axis ($\mathbf{s}_y=\mathbf{q}_y-\mathbf{p}_y, \mathbf{c}_y={\mathbf{s}_y}/{2}$) and sample the $x$ coordinates as 
\begin{align} 
    \mathbf{c}_x&\sim \mathcal{U}\left(\mathbf{p}_x+\delta_x, \mathbf{q}_x-\delta_x\right) \\
    \mathbf{s}_x&\sim \mathcal{U}\left(s_x^{\text{min}}, \min\{\mathbf{c}_x, \mathbf{N}_x - \mathbf{c}_x\}\right)\,,
\end{align}
where we set $s_x^{\text{min}}$ and $\delta_x$ to the 10\% and the 25\% of $\mathbf{q}_x-\mathbf{p}_x$, respectively.
For the $z$ coordinates, since we require the dam to touch the ground, we only sample the center
\begin{equation}
\mathbf{c}_z\sim\mathcal{U}\left(\mathbf{p}_z+\delta_z, \dfrac{\mathbf{q}_z-\mathbf{p}_z}{2}\right)
\end{equation}
and set $\mathbf{s}_z=2(\mathbf{c}_z-\mathbf{p}_z)$, where $\delta_z$ is set to the 25\% of $\mathbf{q}_z-\mathbf{p}_z$.

\begin{table}[t]
    \centering
    \caption{Real-world scale of the \texttt{obstacles} and \texttt{dam-break} datasets. Legend: $\delta x, \delta m, \delta t \rightarrow$ space, mass and time units (respectively). $\nu\rightarrow$ kinematic shear viscosity. $\sigma\rightarrow$ surface tension. $\rho\rightarrow$ density. $g\rightarrow$ gravitational acceleration. $Re_{\text{max}}\rightarrow$ upper bound for the Reynolds number of the simulations.}
    \label{tab:real-world-scale}
    \begin{tabular}{c | c  c}
         \toprule
                    & Real                                  & LBM \\
         \midrule
         $\delta x$ & $7.81\cdot10^{-3}\left[m\right]$      & 1  \\
         $\delta m$ & $4.76\cdot10^{-4}\left[kg\right]$     & 1  \\
         $\delta t$ & $4.45\cdot10^{-4}\left[s\right]$      & 1  \\
         $\nu$      & $2.00\cdot10^{-3}\left[m^2/s\right]$  & $1.46\cdot10^{-2}$ \\
         $\sigma$   & $7.20\cdot10^{-2}\left[kg/s^2\right]$ & $3.00\cdot10^{-4}$ \\
         $\rho$     & $1.00\cdot10^{3}\left[kg/m^3\right]$  & 1 \\
         $g$        & $9.81\left[m/s^2\right]$              & $2.49\cdot10^{-4}$ \\
         $Re_{\text{max}}$ & 10119 & 10119 \\
         \bottomrule
    \end{tabular}
\end{table}


We choose physical properties of the fluid to achieve: a) reasonable space/time scale, b) water-like behavior, c) stability of the simulation. The resulting real world scale, coupled with the corresponding LBM units, is detailed in \Cref{tab:real-world-scale}.
We fix the simulation domain to have characteristic length $L=2[m]$ in all dimensions, and  select density and surface tension values to be identical to those of water. To obtain the remaining quantities, we need to fix the simulation resolution; we choose $\mathbf{N}=(256,256,256)$, which we observed to yield a reasonable tradeoff between quality and efficiency. From the fixed data, we obtain $\delta x=L/\text{max}\{\mathbf{N}\}$, and $\delta m = \rho \cdot \delta x^3$. To determine the timestep, we fix the surface tension in LBM units and obtain $\delta t = \sqrt{\sigma_{\text{real}} / \sigma_{\text{LBM}} \cdot \delta m}$. The value is empirically chosen to obtain a reasonable time scale. In this setting, using the real kinematic shear viscosity of water ($10^{-6}$) would dramatically increase the Reynolds number $Re_{\text{max}}={(\text{min}\{\mathbf{N}\} \cdot \mathbf{u}_{\text{max}})}/{\nu_{\text{LBM}}}$ (where $\mathbf{u}_{\text{max}}=1/\sqrt{3}$ in LBM) of our simulations, yielding turbulent flows which can make the simulation unstable. Therefore, we reduce the viscosity to keep the maximum Reynolds number around $10^4$, which we verified ex-post to yield stable simulations.

\begin{table}[t]
    \centering
    \caption{Generation data for our neural fluid simulation datasets. We aim for an output resolution of 50FPS, therefore we export one frame every 0.02s of simulation (equivalently, 42 LBM steps with the $\delta t$ in \Cref{tab:real-world-scale}). The resulting throughput proves the efficiency of our procedure: allowing for a $10\%$ decay in performance, we could be able to generate a massive 1mln frames dataset (ca. $16\cdot10^{12}$ voxels of data with the same settings) in $\sim$ 45 hours.}
    \label{tab:datasets-info}
    \resizebox{\columnwidth}{!}{
    \begin{tabular}{l | c c}
        \toprule
        Dataset & \texttt{dam-break} & \texttt{obstacles} \\
        \midrule
        Generation time (s) & 15253 & 14654 \\  
        Throughput (FPS) & 6.55 & 6.82 \\
        Sampled scenes & 200 & 400 \\
        $t_{\text{max}}$ (s) & 10.0 & 5.0 \\
        Frame count & 100.000 & 100.000 \\
        Frame time (s) & 0.02 & 0.02 \\
        Size (GB) & 23.2 & 17.5 \\
        Particles & 5000 & 3000 \\
        
        \bottomrule
    \end{tabular}%
    }
\end{table}

We give complete details on the generation of the two datasets in \Cref{tab:datasets-info}.

\subsubsection{Hand-crafted multi-view scenes}\label{sec:multiview}

Additionally, we used our generation procedure to simulate a limited set of hand-crafted scenes, primarily intended for training inverse rendering and surface recovery algorithms. These scenes are not designed to cover a wide spectrum of fluid behaviors but rather serve as challenging test cases to assess the generalizability and accuracy of novel methods. Details about the \texttt{ship} scene's data are summarized in \Cref{fig:nerf-cameras}, and comprehensive information on real-world scale and simulation for each scene is provided in \Cref{tab:nerf-info}.


We simulate all scenes with resolution $\mathbf{N}=(256,256,256)$ and export at 60FPS. 10 viewpoints are sampled in the upper hemisphere via Fibonacci sphere sampling~\cite{keinert:2015:fibonaccisphere}, and each output frame is rendered with resolution $1080\times1080$ twice, once with a water-like material and once with a Lambertian material. 
For each camera, a background-only render is also produced, and the camera parameters for the viewpoints are exported as JSON. 
The data totals 20.32GB.


\begin{figure}[!htpb]
    \centering
    
\begin{tabular}{c c c}
    $t$ & \textbf{DLF} & \textbf{GT}  \\
    0.0s & 
    \begin{minipage}{0.175\textwidth}
    \includegraphics[width=\textwidth,trim=5.5cm 4cm 3.5cm 6cm,clip]{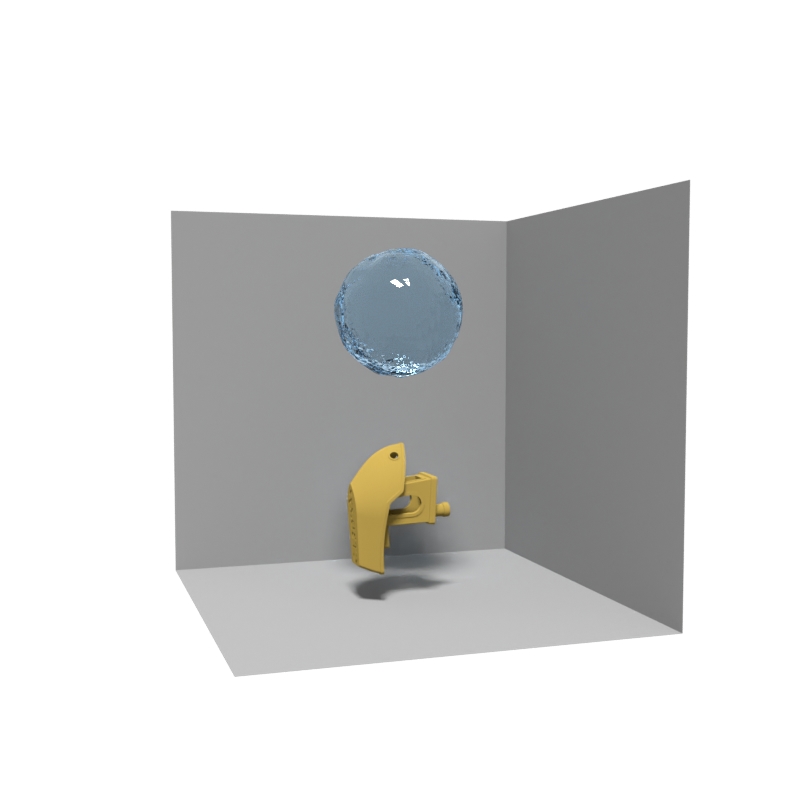}
    \end{minipage} & 
    \begin{minipage}{0.175\textwidth}
    \includegraphics[width=\textwidth,trim=5.5cm 4cm 3.5cm 6cm,clip]{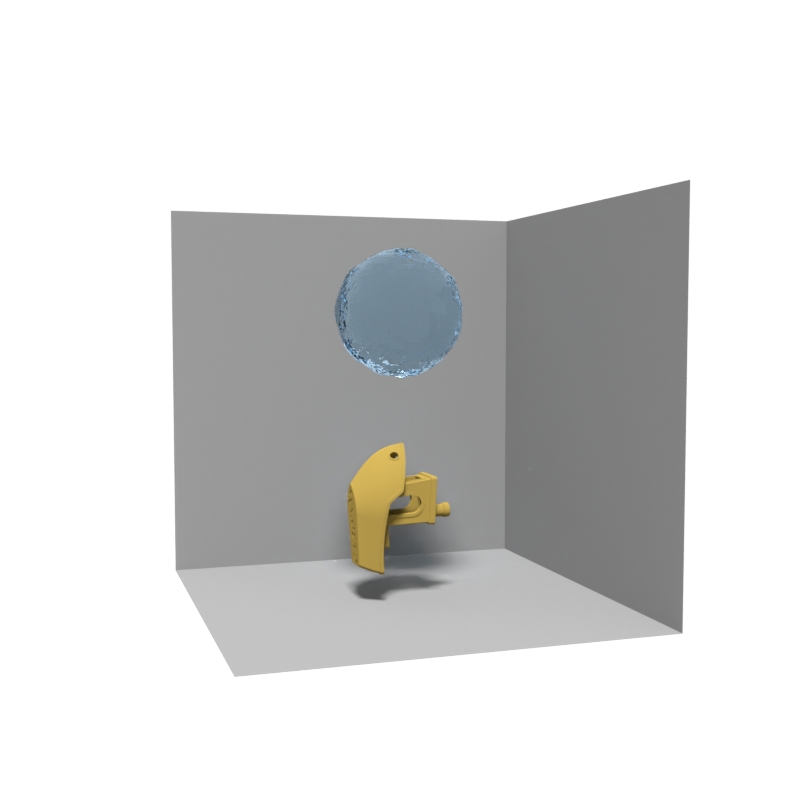}
    \end{minipage} \\
    0.2s & 
    \begin{minipage}{0.175\textwidth}
    \includegraphics[width=\textwidth,trim=5.5cm 4cm 3.5cm 6cm,clip]{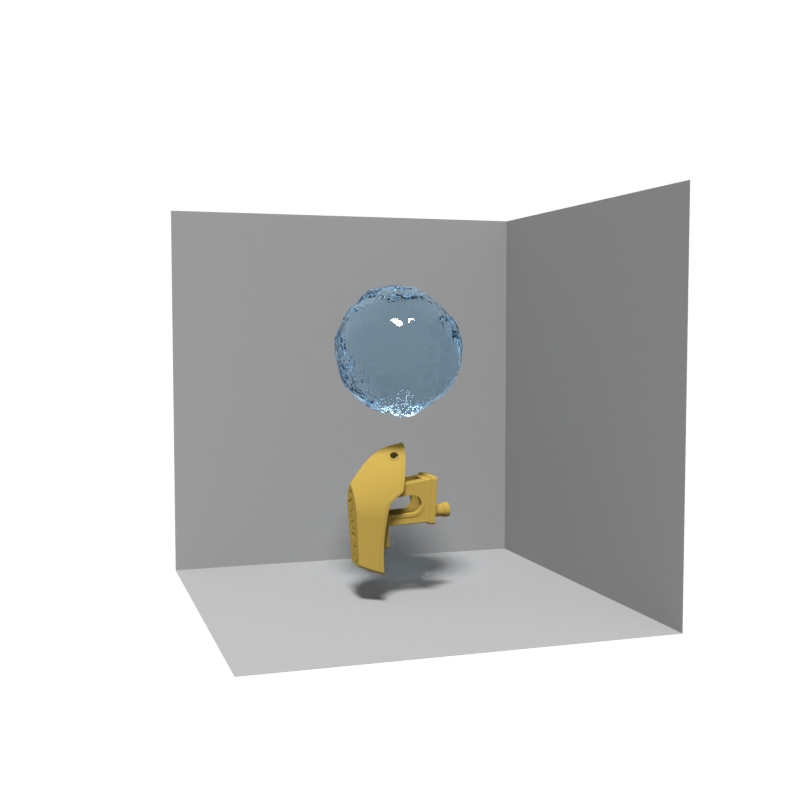}
    \end{minipage} & 
    \begin{minipage}{0.175\textwidth}
    \includegraphics[width=\textwidth,trim=5.5cm 4cm 3.5cm 6cm,clip]{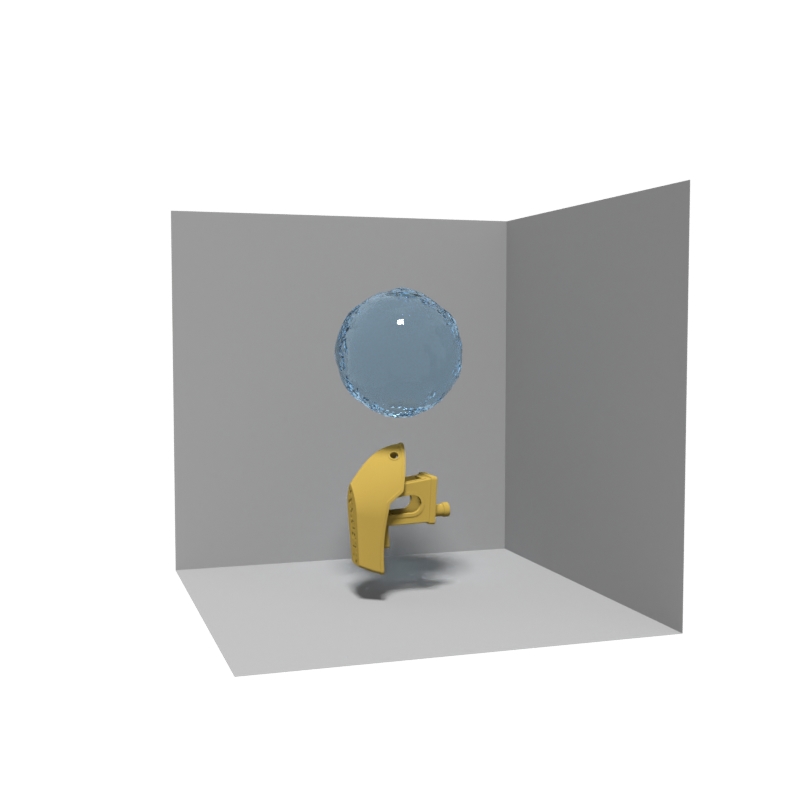}
    \end{minipage} \\
    0.4s & 
    \begin{minipage}{0.175\textwidth}
    \includegraphics[width=\textwidth,trim=5.5cm 4cm 3.5cm 6cm,clip]{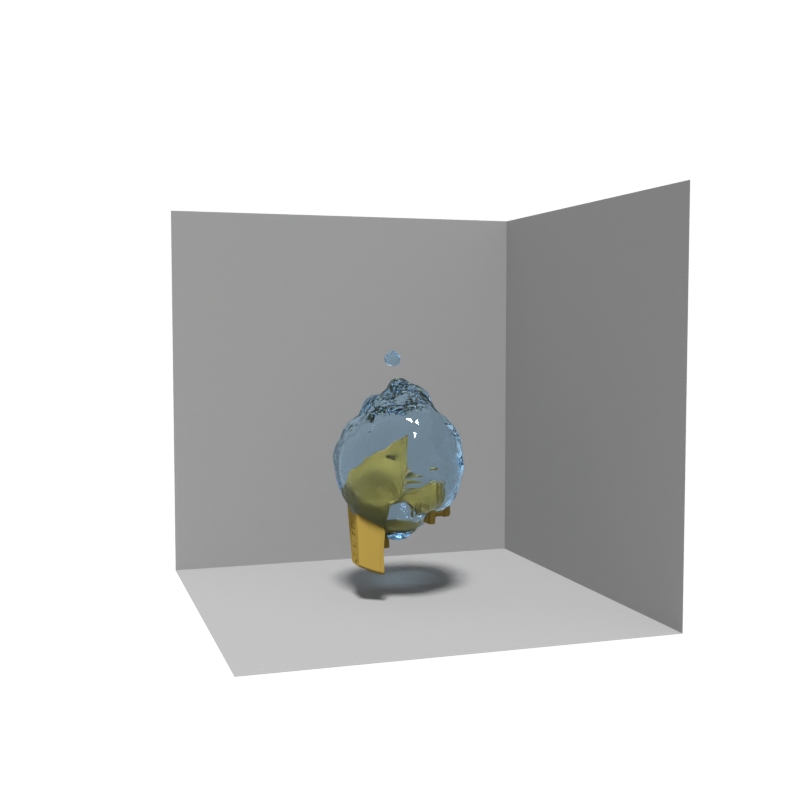}
    \end{minipage} & 
    \begin{minipage}{0.175\textwidth}
    \includegraphics[width=\textwidth,trim=5.5cm 4cm 3.5cm 6cm,clip]{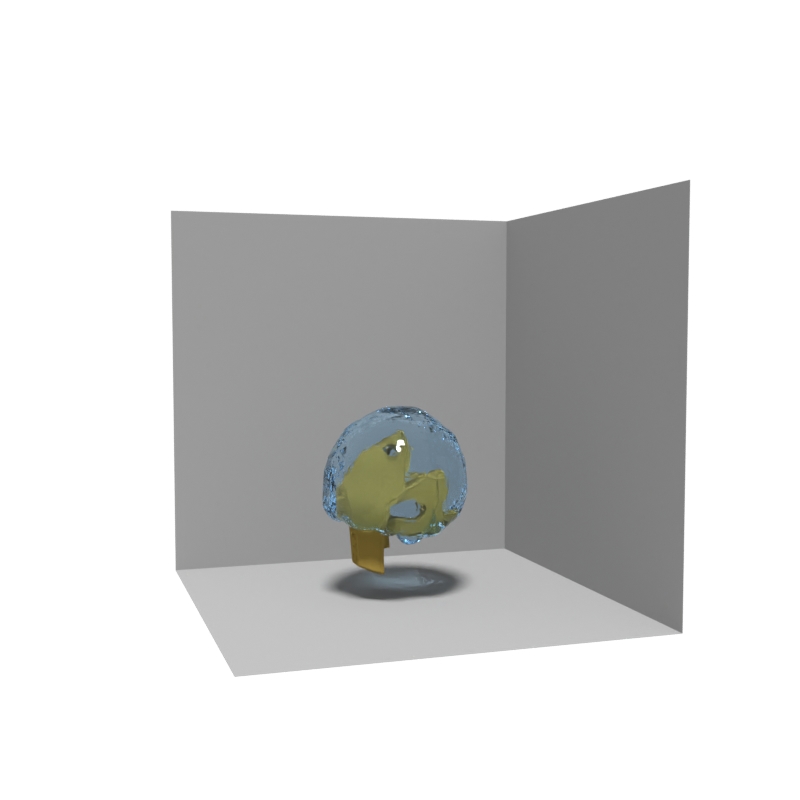}
    \end{minipage} \\
    0.6s & 
    \begin{minipage}{0.175\textwidth}
    \includegraphics[width=\textwidth,trim=5.5cm 4cm 3.5cm 6cm,clip]{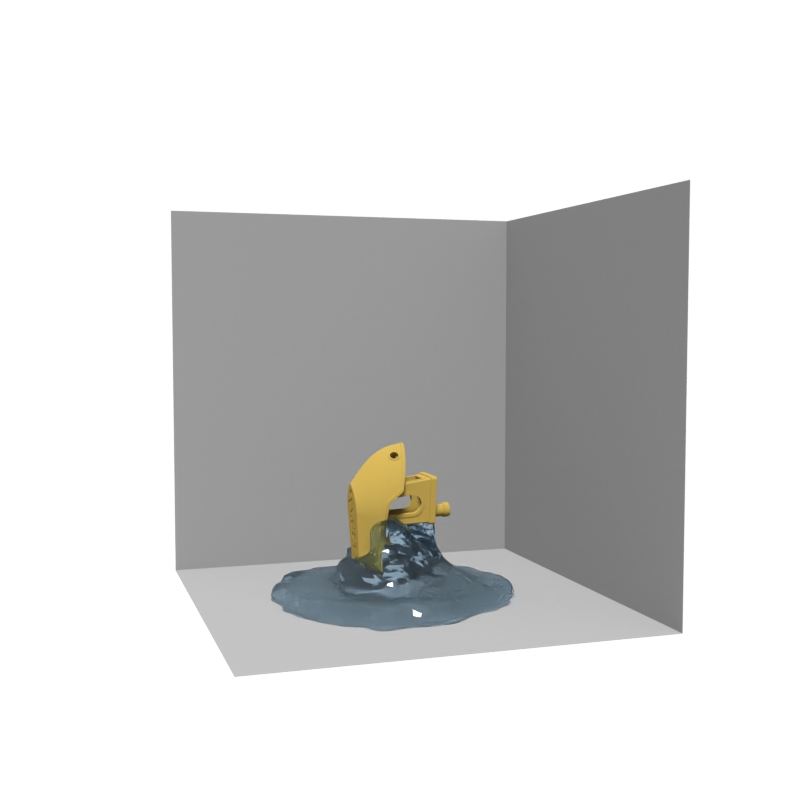}
    \end{minipage} & 
    \begin{minipage}{0.175\textwidth}
    \includegraphics[width=\textwidth,trim=5.5cm 4cm 3.5cm 6cm,clip]{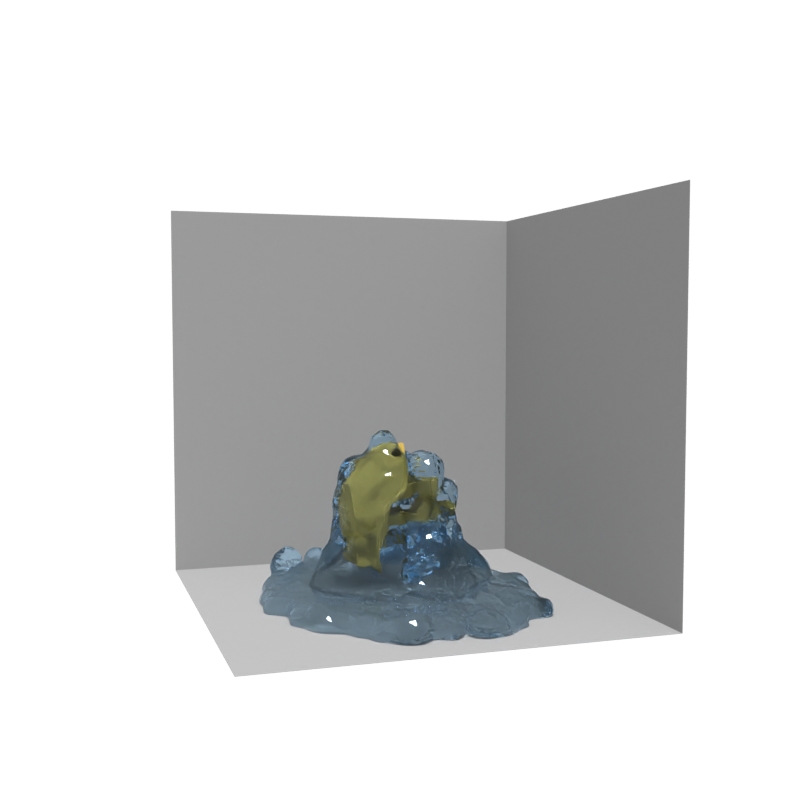}
    \end{minipage} \\
    0.8s & 
    \begin{minipage}{0.175\textwidth}
    \includegraphics[width=\textwidth,trim=5.5cm 4cm 3.5cm 6cm,clip]{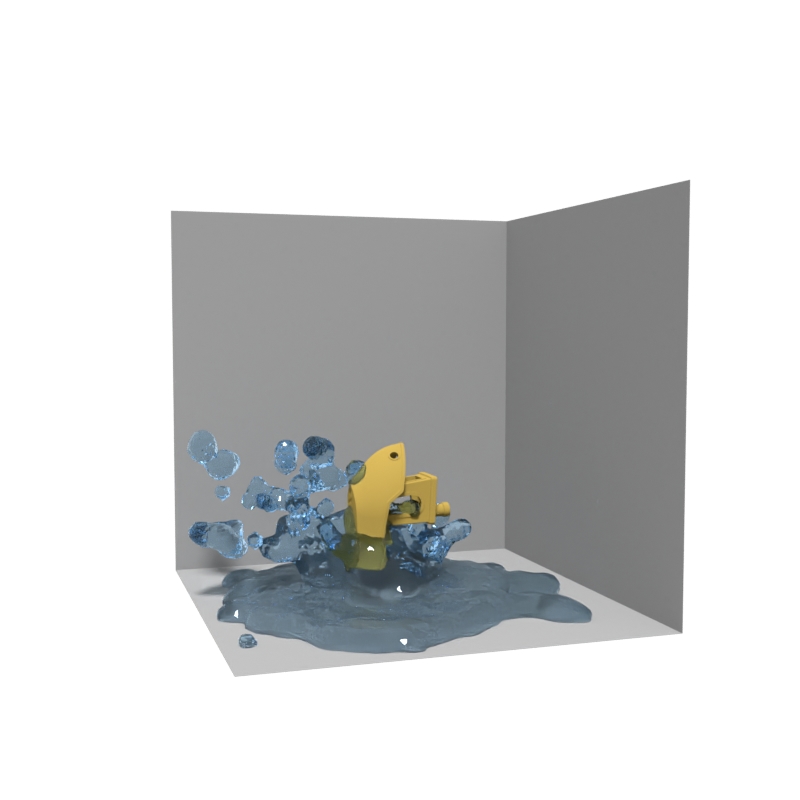}
    \end{minipage} & 
    \begin{minipage}{0.175\textwidth}
    \includegraphics[width=\textwidth,trim=5.5cm 4cm 3.5cm 6cm,clip]{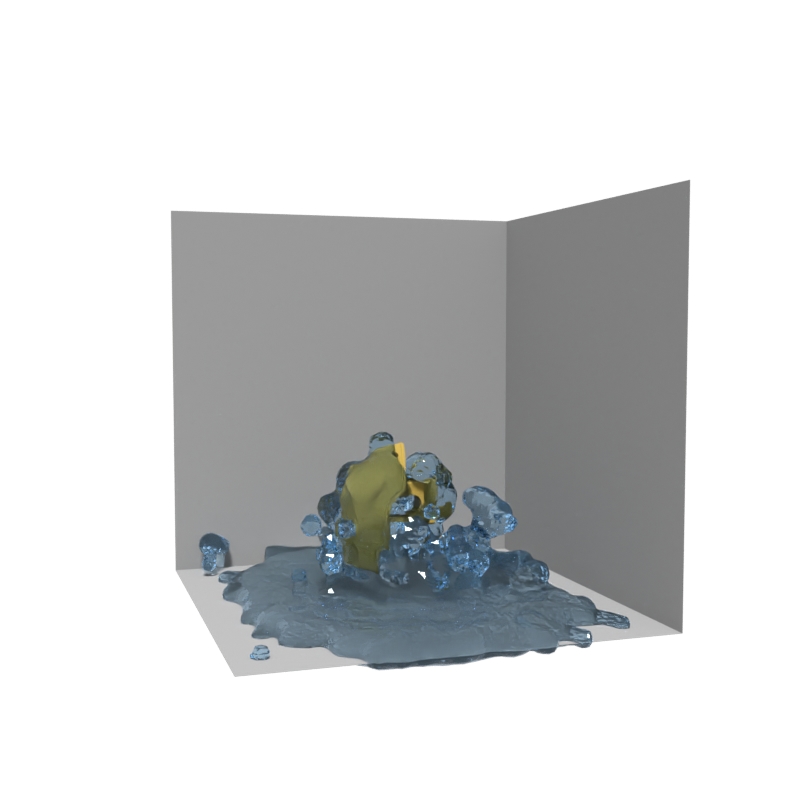}
    \end{minipage} \\
    1.0s & 
    \begin{minipage}{0.175\textwidth}
    \includegraphics[width=\textwidth,trim=5.5cm 4cm 3.5cm 6cm,clip]{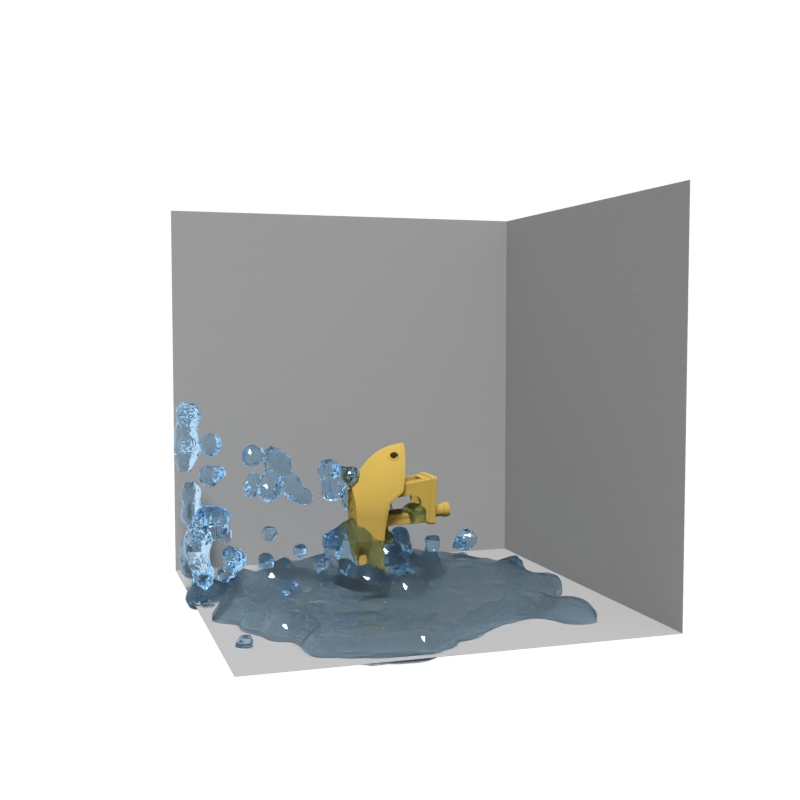}
    \end{minipage} & 
    \begin{minipage}{0.175\textwidth}
    \includegraphics[width=\textwidth,trim=5.5cm 4cm 3.5cm 6cm,clip]{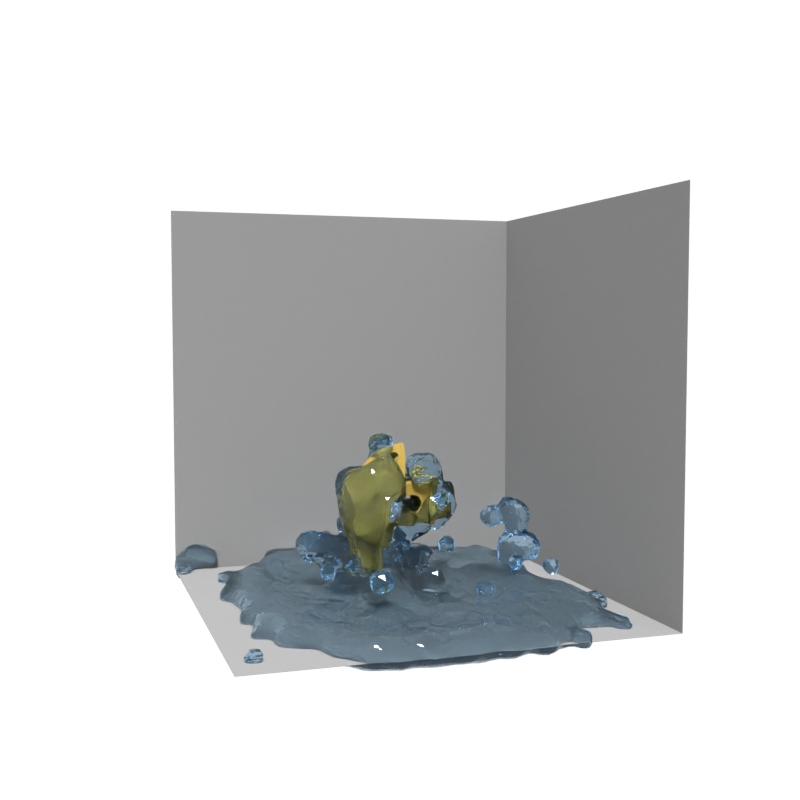}
    \end{minipage} \\
    5.0s & 
    \begin{minipage}{0.175\textwidth}
    \includegraphics[width=\textwidth,trim=5.5cm 4cm 3.5cm 6cm,clip]{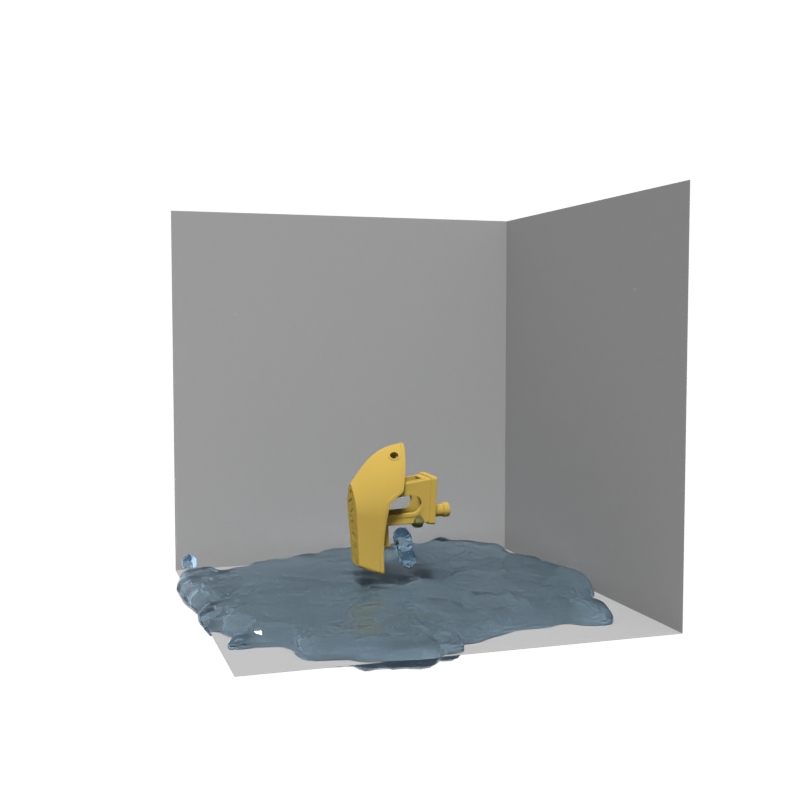}
    \end{minipage} & 
    \begin{minipage}{0.175\textwidth}
    \includegraphics[width=\textwidth,trim=5.5cm 4cm 3.5cm 6cm,clip]{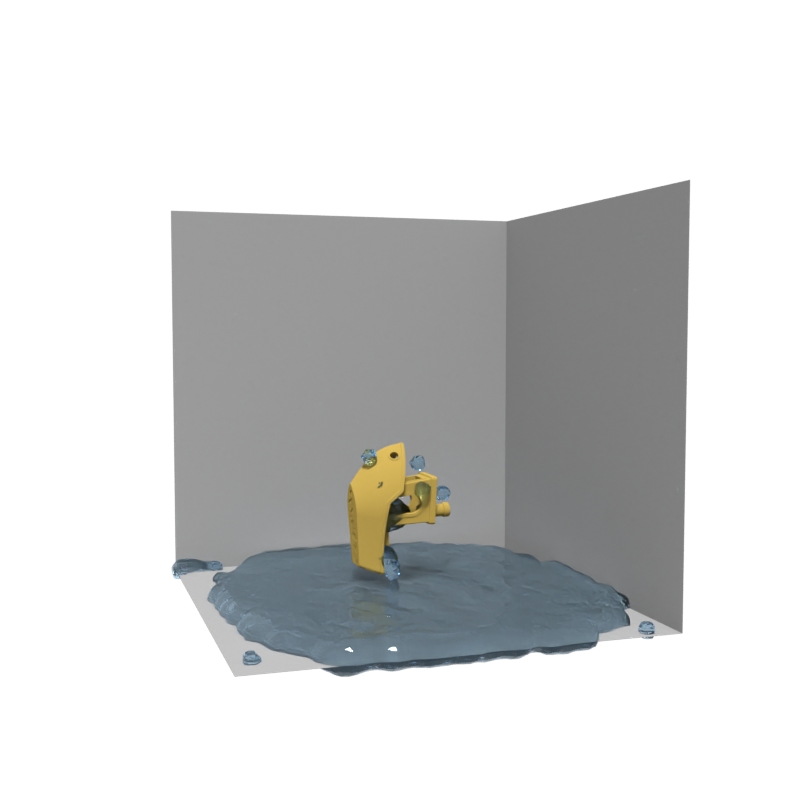}
    \end{minipage} \\
\end{tabular}

    \caption{
    Samples extracted from a 50 frames rollout of a DLF~\cite{Ummenhofer2020Lagrangian} model trained on our \texttt{obstacles} dataset, with unseen initial conditions, compared to the ground truth LBM simulation. 
    }
    \label{fig:dlf-comparison}
\end{figure}


\subsubsection{Copyright information}

Along with our dataset, we redistribute a small set of meshes and HDRI environment textures which we used to make our datasets more interesting. All such content was obtained from \url{polyhaven.com} under CC-0 copyright license (no rights reserved). 

\begin{table}[t]
    \centering
    \caption{Generation time and real-world scale for the scenes in our inverse rendering dataset. $\rho$ and $g$ are as in \Cref{tab:real-world-scale} for all scenes. We fix the reference real world $\sigma=0.072\left[kg/s^2\right]$ and vary the LBM coefficient for surface tension. }
    \label{tab:nerf-info}
    \resizebox{\columnwidth}{!}{
    \begin{tabular}{l | c c c c}
        \toprule
        Scene & \texttt{ball} & \texttt{dam} & \texttt{duck} & \texttt{ship} \\
        \midrule
        $t_{\text{max}}$ (s) & 1.5 & 2.5 & 1.5 & 2.5 \\
        Frame count          & 90  & 150 & 90  & 150 \\
        \midrule
        $L\left[m\right]$               & 2.0                & 2.0                & 2.0                & 2.0 \\      
        $\delta x\left[m\right]$        & $7.81\cdot10^{-3}$ & $7.81\cdot10^{-3}$ & $7.81\cdot10^{-3}$ & $7.81\cdot10^{-3}$ \\
        $\delta m\left[kg\right]$       & $4.76\cdot10^{-4}$ & $4.76\cdot10^{-4}$ & $4.76\cdot10^{-4}$ & $4.76\cdot10^{-4}$ \\
        $\delta t\left[s\right]$        & $8.13\cdot10^{-4}$ & $2.57\cdot10^{-4}$ & $8.13\cdot10^{-4}$ & $8.13\cdot10^{-4}$ \\
        $\nu\left[m^2/s\right]$         & $5\cdot10^{-4}$    & $2\cdot10^{-3}$    & $5\cdot10^{-2}$    & $5\cdot10^{-3}$ \\
        $\sigma\left[\text{LBM}\right]$ & $1\cdot10^{-4}$    & $1\cdot10^{-5}$    & $1\cdot10^{-4}$    & $1\cdot10^{-4}$ \\
        $Re_{\text{max}}$               & 22170              & 17527              & 222                & 2217 \\
        \bottomrule
    \end{tabular}%
    }
\end{table}

\section{Evaluation}

\subsection{Examples of applications}

\subsubsection{Data-driven Fluid Simulation}
In this section, we test the suitability of the datasets described in \Cref{sec:benchmarks} to train a data-driven model.

\paragraph*{Model}
\citet{Ummenhofer2020Lagrangian} aims to understand fluid mechanics by studying particle motion. In particular, a continuous convolutional network is fed with a collection of particles, each paired with its features. Each particle is associated with a feature vector that is a constant scalar of 1, paired with the particle's velocity, denoted by $\mathbf{v}$, and its viscosity, $\nu$. Therefore, at any timestep $n$, a particle $p_i^n$ is represented by the tuple $\left(\mathbf{x}_i^n,\left[1, \mathbf{v}_i^n, \nu_i\right]\right)$.
To compute the intermediate velocities and positions and integrate external force information for the network, the velocity is listed as an input feature. Using Heun's method, the intermediate velocities $\mathbf{v}_i^{n *}$ and positions $\mathbf{x}_i^{n *}$ commencing from timestep $n$ are  then computed as:
\begin{align}\label{eq:heun1}
    & \mathbf{v}_i^{n *}=\mathbf{v}_i^n+\Delta t \mathbf{a}_{\mathrm{ext}} \\
    & \mathbf{x}_i^{n *}=\mathbf{x}_i^n+\Delta t \frac{\mathbf{v}_i^n+\mathbf{v}_i^{n *}}{2}\label{eq:heun2}
\end{align}
where $\mathbf{a}_{\text {ext }}$ represents an acceleration vector, enabling the application of external forces like fluid control or gravity. These intermediate values are devoid of any particle or scene interactions; such interactions are incorporated using the ConvNet. For the network to manage collisions, another group of static particles $s_j$ are introduced. These are sampled along scene boundaries and paired with normals $\mathbf{n}_j$ as their feature vectors, expressed as $s_j=\left(\mathbf{x}_j,\left[\mathbf{n}_j\right]\right)$.
The network performs the function:
\begin{equation}\label{eq:conv}
\left[\Delta \mathbf{x}_1, \ldots, \Delta \mathbf{x}_N\right]=\operatorname{ConvNet}\left(\left\{p_i^{n *}\right\}_{i=1}^{N},\left\{s_i\right\}_{i=1}^{M}\right),
\end{equation}
employing convolutions to merge features from both sets of particles. Here, $\Delta \mathrm{x}$ serves as a position correction, factoring in all particle interactions, including collisions with the scene boundary. The correction is finally used to update positions and velocities for timestep $n+1$ with:
\begin{align}\label{eq:dlf-update1}
    \mathbf{x}_i^{n+1} & =\mathbf{x}_i^{n *}+\Delta \mathbf{x}_i \\ 
    \mathbf{v}_i^{n+1} & =\frac{\mathbf{x}_i^{n+1}-\mathbf{x}_i^n}{\Delta t} \label{eq:dlf-update2}
\end{align}
We refer to this model as \textit{Deep Lagrangian Fluids} (DLF).

\paragraph*{Experiment and results} 
We train DLF over \texttt{obstacles}, split as train and validation sets with a 90:10 ratio, which amounts to 360 training simulations (tot. 90k frames) and 40 validation simulations (tot. 10k frames). 
\Cref{fig:dlf-comparison} shows a qualitative comparison of the model prediction vs.~the ground truth simulation, after training DLF to convergence (50,000 total iterations), given an initial state from the validation set. The model is able to consistently predict realistic dynamics throughout the simulation, and reach a stable state coherent with the one shown in the data, proving our data to be suitable for this application; the only shortcoming of the learned model is its inability to properly capture viscous behavior, probably due to the viscosity not being explicitly modeled in its formulation (\cref{eq:heun1,eq:heun2,eq:conv,eq:dlf-update1,eq:dlf-update2}).

\begin{figure}[t]
    \centering
    \includegraphics[width=0.3\columnwidth]{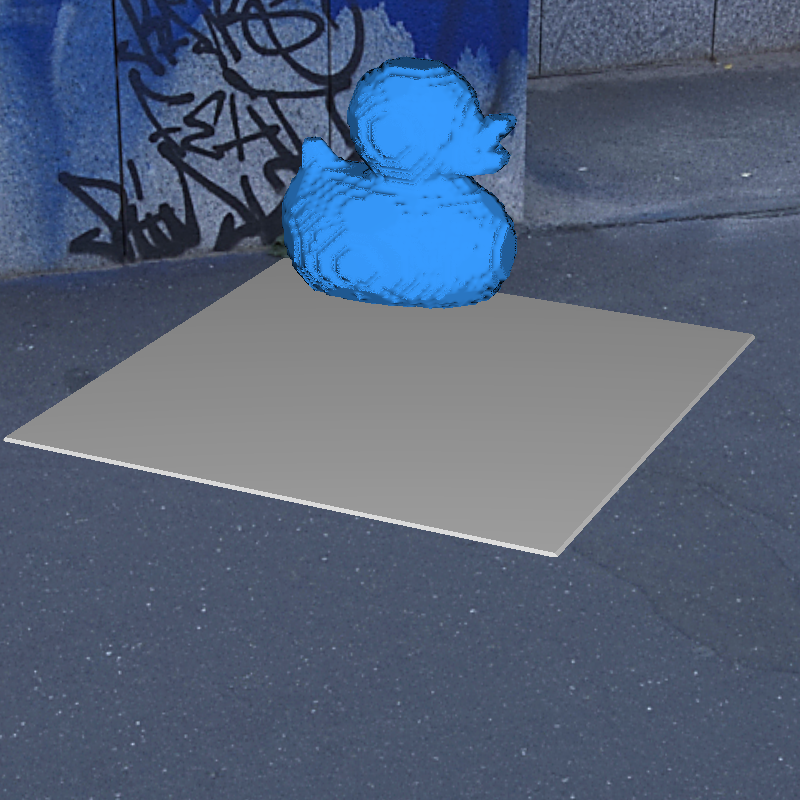} 
    \hfill
    \includegraphics[width=0.3\columnwidth]{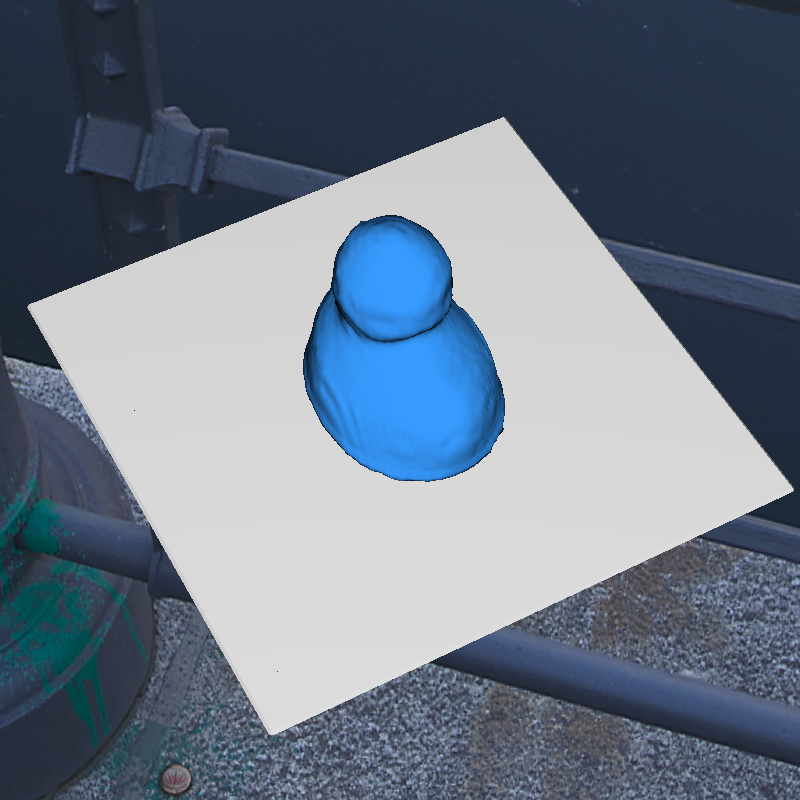}
    \hfill
    \includegraphics[width=0.3\columnwidth]{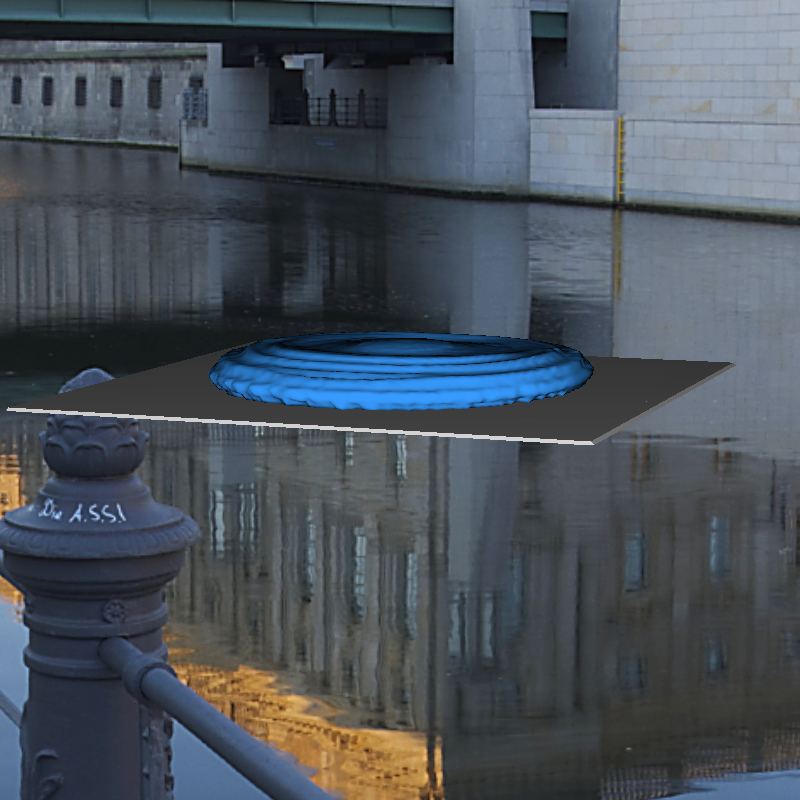}
    
    \includegraphics[width=0.3\columnwidth]{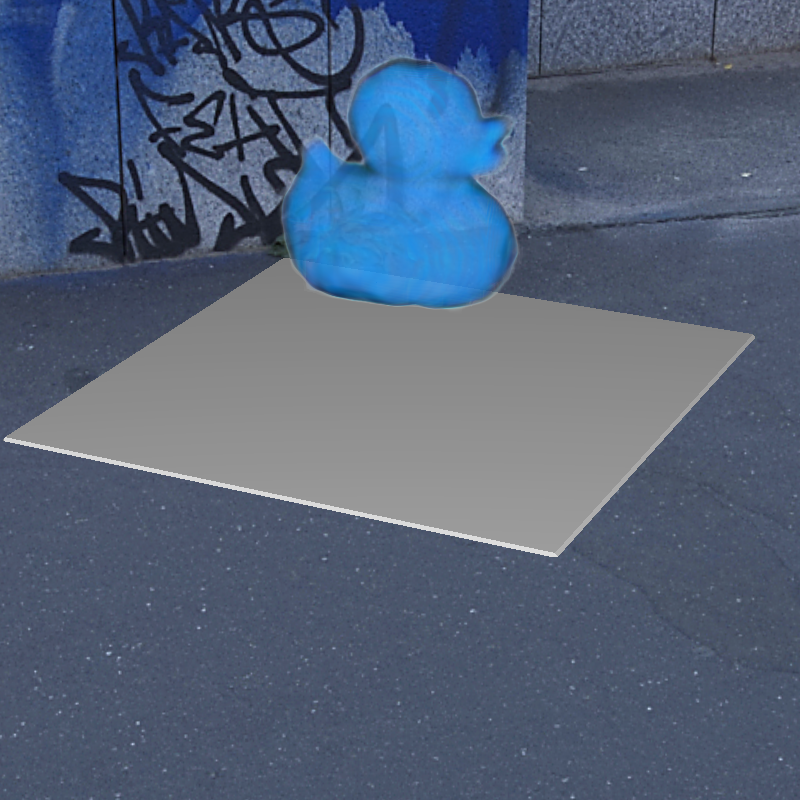} 
    \hfill
    \includegraphics[width=0.3\columnwidth]{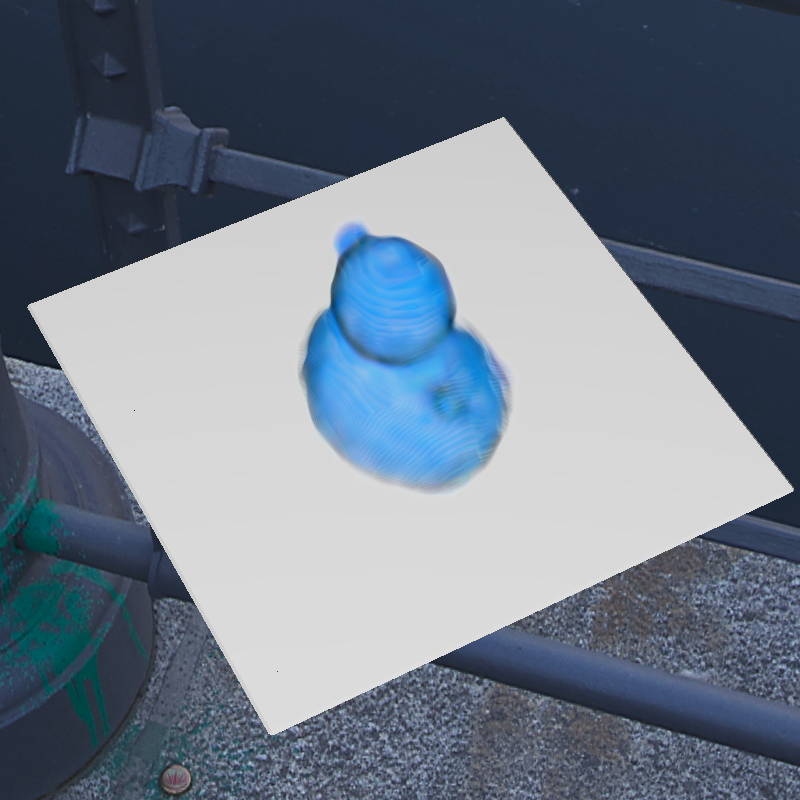}
    \hfill
    \includegraphics[width=0.3\columnwidth]{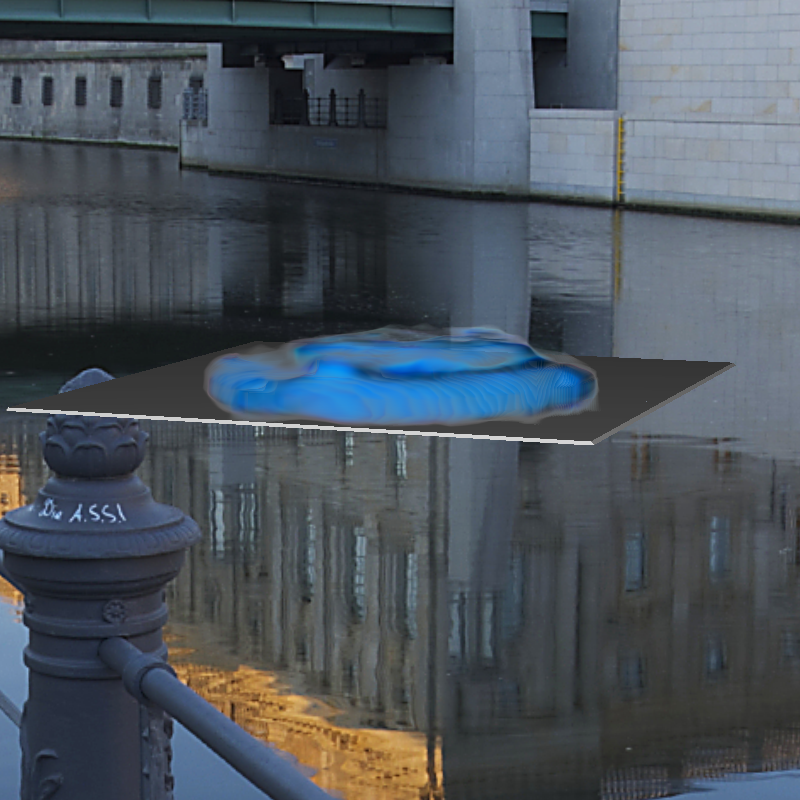}
    \caption{Top: a subset of the views in our \texttt{duck} scene training data. Bottom: PAC-NeRF \cite{li2023pacnerf} reconstruction and rendering for the same (unseen) views (the background is not learned by the model, so it was manually composited for this visualization).}
    \label{fig:nerf-comparison}
\end{figure}

\begin{figure}[h]
    \centering
    \includegraphics[width=\columnwidth]{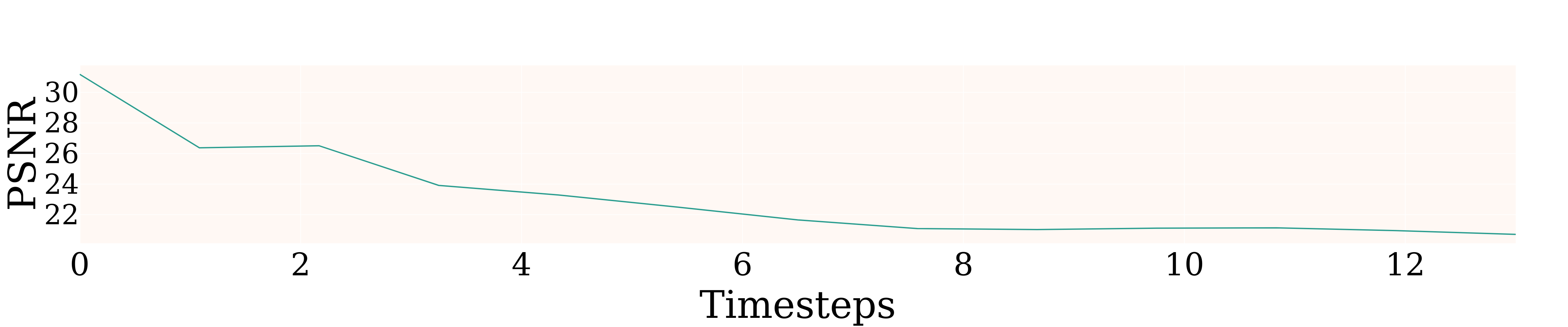} 
    \includegraphics[width=\columnwidth]{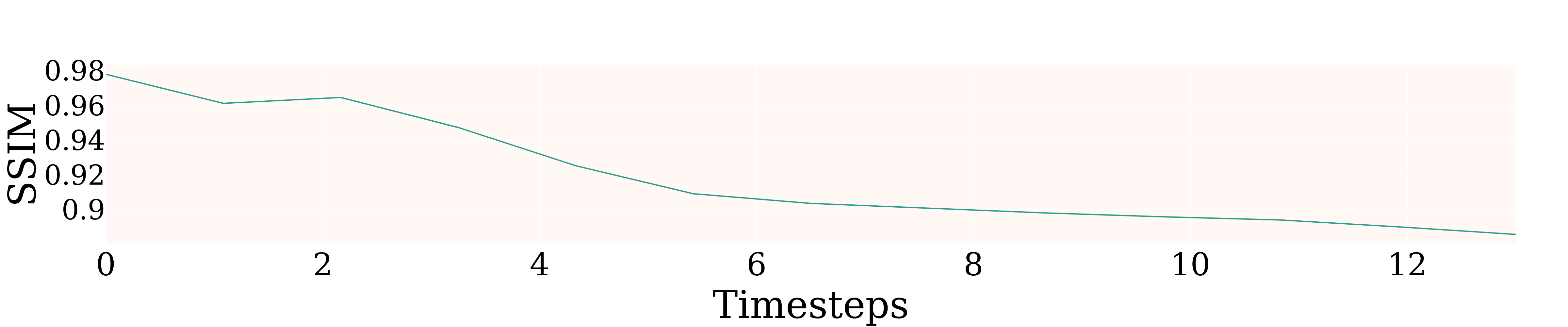}
    \includegraphics[width=\columnwidth]{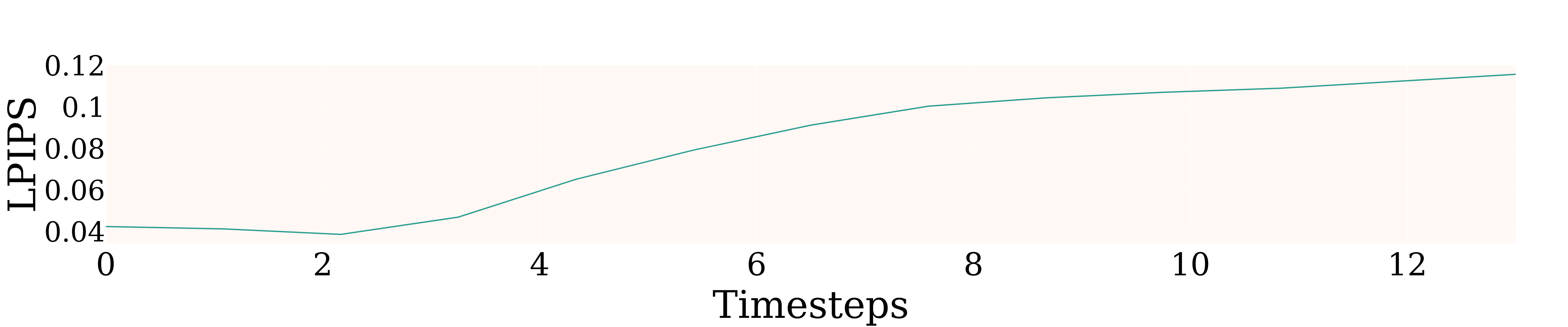}
    \includegraphics[width=\columnwidth]{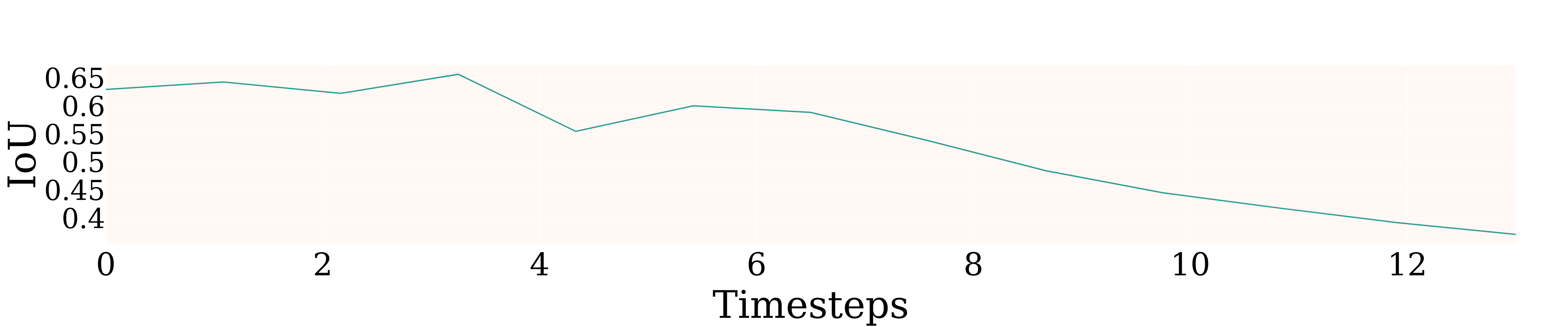}
    \caption{PAC-NeRF evaluation on \texttt{duck}. Both rendering quality and IoU over occupancy grids degrade as the simulation progresses because the neural radiance field reconstruction gets increasingly constrained by the physics-based losses. Note that the image metrics are computed only on the masked foreground object.}
    \label{fig:nerf-metrics}
\end{figure}


\subsubsection{Inverse Rendering Fluid Simulations}

Our method allows generating multi-view video data for inverse rendering, consisting of image-pose pairs over time.  By defining a $\Delta t$ and a set of randomly sampled cameras on the upper hemisphere (see Figure~\ref{fig:nerf-cameras}), we render the scene from these viewpoints during simulation. While our data can be used to train various NeRF-based models, both static and dynamic, we tested them on PAC-NeRF \cite{li2023pacnerf}. PAC-NeRF recovers explicit geometric representations and physical properties of dynamic objects in scenes by combining neural scene representations with differentiable physics engines for continuum materials.

A dynamic NeRF comprises time-dependent volume density field $\sigma(\mathbf{x}, t)$ and a time-and-view-dependent appearance (color) field $\mathbf{c}(\mathbf{x}, \omega, t)$ for each point $\mathbf{x} \in \mathbb{R}^3$, and directions $\omega=(\theta, \phi) \in \mathbb{S}^2$ (spherical coordinates). The appearance $\mathbf{C}(\mathbf{r}, t)$ of a pixel specified by ray direction $\mathbf{C}(\mathbf{r}, t)$ $\left(s \in\left[s_{\min }, s_{\max }\right]\right)$ is obtained by differentiable volume rendering \cite{mildenhall2020nerf}:
\begin{equation}
\begin{split}
\mathbf{C}(\mathbf{r}, t) & =  \mathbf{c}_{b g} T\left(s_f, t\right) + \\ 
& \int_{s_n}^{s_f} T(s, t) \sigma(\mathbf{r}(s), t) \mathbf{c}(\mathbf{r}(s), \omega, t) d s
\end{split}
\end{equation}
\begin{equation}
    T(s, t)=\exp \left(-\int_{s_n}^s \sigma(\mathbf{r}(\bar{s}), t) d \bar{s}\right)
\end{equation}
The dynamic NeRF can therefore be trained to minimize the rendering loss:
\begin{equation}
\mathcal{L}_{\text {render }}=\frac{1}{N} \sum_{i=0}^{N-1} \frac{1}{|\mathcal{R}|} \sum_{\mathbf{r} \in \mathcal{R}}\left\|\mathbf{C}\left(\mathbf{r}, t_i\right)-\hat{\mathbf{C}}\left(\mathbf{r}, t_i\right)\right\|^2
\end{equation}
where $N$ is the number of frames of videos, $\hat{\mathbf{C}}\left(\mathbf{r}, t_i\right)$ is the ground truth color observation.

PAC-NeRF first initializes an Eulerian voxel field over the first frame of the sequence. It then uses a grid-to-particle conversion method to obtain a Lagrangian particle field; this is advected, enforcing that the appearance and volume density fields admit conservation laws characterized by the velocity field $\mathbf{v}$ of the underlying physical system:
\begin{equation}
\frac{D \sigma}{D t}=0, \quad \frac{D \mathbf{c}}{D t}=0
\end{equation}
with $\frac{D \phi}{D t}=\frac{\partial \phi}{\partial t}+\mathbf{v} \cdot \nabla \phi$ being the material derivative of an arbitrary time-dependent field $\phi(\mathbf{x}, t)$. Moreover, the velocity field must obey momentum conservation for continuum materials:
\begin{equation}
\rho \frac{D \mathbf{v}}{D t}=\nabla \cdot \boldsymbol{T}+\rho \mathbf{g}
\end{equation}
where $\rho$ is the physical density field, $\boldsymbol{T}$ is the internal Cauchy stress tensor, and $\mathbf{g}$ is the acceleration due to gravity and it is evolved using a differentiable Material Point Method (MPM) \cite{hu2018moving}.
The advected field is then mapped back to the Eulerian domain using the particle-to-grid conversion and is used for collision handling and neural rendering. We refer to the original paper for more technical details.


\paragraph*{Experiment and results} 
We trained PAC-NeRF on \texttt{duck} by subsampling a 1$s$ simulation of 1800 frames down to 13 frames with a $\Delta t$ of $\approx 0.077 s$.
Figure~\ref{fig:nerf-comparison} visually confirms that our model generates accurate renders of simulations, validating our multi-view data generation for use in inverse rendering techniques within fluid dynamics.
PAC-NeRF applies stricter advection constraints to the NeRF model compared to other 4D NeRF methods. This enhances scene dynamics recognition and yields smoother, more realistic inter-frame reconstructions. However, it may compromise the supervised frame reconstruction quality over time, as depicted in Figure~\ref{fig:nerf-metrics}, unlike other 4D NeRF methods which maintain stable reconstruction quality but may result in less physically accurate interpolations between simulation steps.


\subsection{Generation performance}\label{sec:generationperformance}
%

We assess our generation tool's efficiency, demonstrating that its GPU implementation allows rapid dataset creation. This speed negates the need to share bulky data, as users can replicate any dataset quickly using just the JSON configuration with our tool.

%

We provide data about the generation of our \texttt{obstacles} and \texttt{dam-break} datasets in \Cref{tab:datasets-info}. All scenes are rendered from a single viewpoint at low resolution ($800\times 800$), to offer a visual support to the generated data. Our procedure reaches a throughput of $\sim$6.5 frames per second, allowing to generate 100,000 frames of data in $\sim$ 4 hours. We further analyze our procedure's runtime and memory occupancy by collecting performance data over multiple runs, changing the simulation resolution and the rendering resolution. The results are reported in \Cref{tab:resolutions}: exploiting GPU programming massively benefits both the simulation and rendering steps, so much so that rendering 150 frames even in 4K only accounts for 28\% of the total computation time. Furthermore, this motivates our choice of sparse format for storing density fields: while 4GB per scene may seem a lot at simulation resolution $512$, the dense counterpart would require about 80GB per scene ($150\cdot512^3$ floats), \ie, we only require 5\% of the disk memory.

All generations were run on medium/high-end consumer hardware, to further solidify the claim that our data can be re-generated locally, without the need for sharing entire datasets. Our machine runs Windows 11 over 32GB of DDR4 RAM, an intel core i7 12700K CPU (3.6GHz, 12 physical processors, 20 logical processors), and a NVIDIA RTX4070Ti GPU (7680 CUDA cores). The data was generated on an SSD drive to minimize disk latency.

\begin{table}[t]
    \centering
    \caption{
    Analysis of resource consumption in terms of CPU memory, GPU memory, scene size and time, varying rendering and simulation resolution. Values are averaged over 20 repeated runs over the \texttt{dam} scene from the multi-view dataset. Rendering is disabled while gathering the simulation data. Geometry is exported in sparse density format. 
    }
    \label{tab:resolutions}
    \resizebox{\columnwidth}{!}{
    \begin{tabular}{lccccc}
        \toprule
        Rendering resolutions & $800 \times 800$ & FHD & 2K & 4K \\
        \midrule
        CPU Memory (MB)      & 374.4 & 374.4 & 374.4 & 374.4  \\
        GPU Memory (MB)   & 1242.4 & 1244.4 & 1246.4 & 1248.4\\
        Scene size (MB)   & 780.35 & 861.44 & 912.94 & 983.42\\
        Time (s) & 16.97 & 20.46 & 21.13 & 22.48 \\
        \toprule
        Simulation resolutions & $64^3$ & $128^3$ & $256^3$ & $512^3$\\
        \midrule
        CPU Memory (MB)      & 5.6 & 46.8 & 374.4 & 2995.2 \\
        GPU Memory (MB)   & 148.6 & 270.8 & 1242.4 & 9015.2 \\
        Scene size (MB)   & 11.12 & 83.12 & 653.2 & 4110.02 \\
        Time (s) & 5.59 & 7.82 & 16.05 & 112.8  \\
        \bottomrule
    \end{tabular}%
    }
\end{table}

\section{Conclusions}\label{sec:conclusion}

Our work introduces an efficient method for generating multimodal fluid simulation data and three benchmark datasets for research. We detail the generation process and software tool interface, achieving a generation speed of 6.5FPS on consumer-grade hardware, facilitating the creation of large-scale datasets efficiently. We demonstrate the utility of our approach by using our datasets to train neural models for data-driven fluid simulation and fluid inverse rendering, validating its value to the research community.



\section{Impact Statement}

This paper presents work whose goal is to advance the field of Machine Learning, specifically the application of data-driven models towards the understanding of complex natural phenomena such as fluid dynamics. There are many potential societal consequences of our work, none which we feel must be specifically highlighted here.

\bibliography{example_paper}
\bibliographystyle{icml2024}




\end{document}